%% file: main.tex
\documentclass[conference]{IEEEtran}

\usepackage{amsmath,amsfonts,amssymb}
\usepackage{booktabs}
\usepackage{graphicx} 
\usepackage{siunitx}
\usepackage{xurl}
\usepackage{tabularx}
\usepackage{algorithm}
\usepackage{algorithmic}

\newcolumntype{L}[1]{>{\raggedright\arraybackslash}p{#1}}
\newcolumntype{C}[1]{>{\centering\arraybackslash}p{#1}}
\newcolumntype{R}[1]{>{\raggedleft\arraybackslash}p{#1}}

\usepackage{array} 
\usepackage{comment}

\title{Hybrid Temporal Differential Consistency Autoencoder for Efficient and Sustainable Anomaly Detection in Cyber-Physical Systems}

\author{

\IEEEauthorblockN{Michael Somma\IEEEauthorrefmark{1}\IEEEauthorrefmark{2}}

\IEEEauthorblockA{
\IEEEauthorrefmark{1}JOANNEUM RESEARCH Forschungsgesellschaft mbH,\\ 
DIGITAL – Institute for Digital Technologies,\\
Steyrergasse 17, Graz, Austria, 8010\\
\IEEEauthorrefmark{2} TU Graz, Institute for Technical Informatics\\
Inffeldgasse 16/I, 8010 Graz \\
Email: \texttt{michael.somma@joanneum.at} 
}
}

\begin{document}

\maketitle

\begin{abstract}
Cyberattacks on critical infrastructure, particularly water distribution systems, have increased due to rapid digitalization and the integration of IoT devices and industrial control systems (ICS). These cyber-physical systems (CPS) introduce new vulnerabilities, requiring robust and automated intrusion detection systems (IDS) to mitigate potential threats. This study addresses key challenges in anomaly detection by leveraging time correlations in sensor data, integrating physical principles into machine learning models, and optimizing computational efficiency for edge applications. We build upon the concept of \textit{temporal differential consistency (TDC) loss} to capture the dynamics of the system, ensuring meaningful relationships between dynamic states \cite{somma_anomaly_2025}. Expanding on this foundation, we propose a hybrid autoencoder-based approach, referred to as \textit{hybrid TDC-AE}, which extends TDC by incorporating both \textit{deterministic} nodes and conventional \textit{statistical} nodes. This hybrid structure enables the model to account for non-deterministic processes. Our approach achieves state-of-the-art classification performance while improving \textit{time to detect anomalies} by 3\%, outperforming the BATADAL challenge leader without requiring domain-specific knowledge, making it broadly applicable. Additionally, it maintains the computational efficiency of conventional autoencoders while reducing the number of fully connected layers, resulting in a more sustainable and efficient solution. The method demonstrates how leveraging \textit{physics-inspired consistency principles} enhances anomaly detection and strengthens the resilience of cyber-physical systems.
\end{abstract}

\vspace{0.2cm}

\section{Introduction}

Recent years have seen a marked increase in cyberattacks targeting critical infrastructure, intensified by the rapid digitalization of these systems. Previously focused on IT network security, the cybersecurity domain is expanding to encompass essential services and critical infrastructures such as water distribution, energy, and transportation \cite{tuptuk_systematic_2021}. Water distribution networks are evolving into cyber-physical systems (CPS) that blend physical operations with digital control via IoT devices and industrial control systems (ICS). This integration exposes new vulnerabilities, increasing the risk of cyber threats that could compromise both safety and security \cite{tuptuk_systematic_2021}. Highlighting the growing urgency of robust cyber defenses, recent incidents have underscored the increasing vulnerability of water systems to cyberattacks.
Incidents in the United States in 2021 exposed weaknesses in outdated access controls \cite{ikeda_two_2021,magill_us_2021}. In Texas, hackers gained access to the SCADA systems of multiple water plants, causing operational disruptions \cite{bajak_hackers_2023}, while in the UK, cybercriminals infiltrated water utilities, compromising login credentials and exfiltrating sensitive data \cite{smith2022southstaffs}. Similarly, a ransomware attack on Veolia North America in 2024 forced systems offline, disrupting customer services \cite{wolfe2024veolia}. 
So far, these attacks have been contained primarily through extensive manual intervention by highly skilled professionals. However, with the increasing frequency of cyber incidents and ongoing demographic shifts leading to potential workforce shortages, reliance on manual responses is becoming unsustainable. This growing challenge underscores the urgent need for automated Intrusion Detection Systems (IDS) to enhance resilience and ensure proactive detection and mitigation of cyber threats to critical infrastructure \cite{alqudhaibi_predicting_2023,alshamrani_survey_2019,stojanovic_apt_2020-1}. \par
This study addresses three key challenges in anomaly detection for CPS, focusing on water distribution systems. The first is leveraging strong time correlations in time series data from physical processes. The second involves enhancing machine learning models by directly integrating physical principles. The third focuses on improving model efficiency for edge applications by maintaining high detection performance while reducing computational demands.
\section{Related Work}
This section follows the problem formulation introduced by \cite{somma_anomaly_2025}, which identifies two central challenges in anomaly detection for complex dynamical systems:  effective time series modeling and the integration of physical principles into machine learning architectures.
 LSTMs (Long Short-Term Memory networks) have proven to be highly effective in capturing temporal dependencies in time series data \cite{siami-namini_comparison_2018,siami-namini_performance_2019}. Recently, there has been growing interest in applying LSTMs within autoencoders to enhance the detection of anomalies in CPS \cite{wei_lstm-autoencoder-based_2023,longari_cannolo_2021,nguyen_forecasting_2021}. While LSTMs are adept at capturing temporal structures, their complexity and resource-intensive nature do not align with our sustainability goals, as they often result in models that are computationally heavy and energy-intensive, and they are not always applicable for edge applications.
Physics-informed neural networks (PINNs) have shown great promise in embedding physical laws directly into neural networks, improving generalization and interpretability in physics-based applications \cite{cai_physics-informed_2021,raissi_physics-informed_2019,raissi_hidden_2020}. However, applying PINNs in complex systems like water distribution networks is particularly challenging. These systems are governed by numerous interdependent physical laws, and remodeling the entire system to integrate these laws into the neural network would require significant computational resources and extensive domain expertise, making it impractical for real-world applications \cite{karniadakis_physics-informed_2021,wang_when_2022}. \par
Recent work \cite{somma_anomaly_2025}
has proposed a system-theoretic approach inspired by classical embedding theory to analyze the dimensionality of a system’s latent representation. This method introduces physics-inspired consistency principles, which approximate the causal mechanisms governing system dynamics without explicitly enforcing physical laws. The underlying idea is that complex systems in a stable regime exhibit predictable behavior, allowing for well-approximated lower-dimensional embeddings. In contrast, anomalies disrupt these stable relationships, introducing additional complexity into the system.  
To incorporate temporal dynamics into the training process, the proposed TDC-AE (Temporal Differential Consistency Autoencoder) employs a \textbf{Temporal Differential Consistency (TDC) loss}. This loss function enforces consistency between the approximated time derivative of static latent variables and the corresponding dynamic latent variables, improving the model’s ability to detect deviations from normal behavior. By integrating lightweight physics-inspired constraints, this method provides an efficient alternative to traditional physics-informed approaches, making it well-suited for real-time anomaly detection in CPS with constrained computational resources.  

Sustainability is a key concern in designing machine learning models for CPS. Training and deployment can be resource-intensive, increasing energy use and carbon emissions. As infrastructure systems such as water distribution prioritize sustainability, traditional methods, although effective in detection, often lack energy efficiency and are less suitable for real-time, large-scale applications. This challenge of sustainability was addressed in previous research \cite{somma_edge-based_2024}, where edge computing was leveraged to achieve substantial gains in energy efficiency for anomaly detection systems. Building on this, we now focus on refining the internal architecture of the model to further enhance detection performance and maximize energy efficiency.

\section{Methodology}

Our approach builds upon two key ideas: the benefits of \textbf{edge computing} for anomaly detection \cite{somma_edge-based_2024} and the \textbf{temporal differential consistency (TDC) loss} framework \cite{somma_anomaly_2025}. While these approaches provide strong foundations, they also present limitations. Edge computing alone does not inherently address the challenge of finding an embedding that accurately captures the system dynamics. While TDC aims to enforce meaningful relationships between static and dynamic states in complex dynamical systems, we hypothesize that a diverse range of systems may require a more refined representation that accounts for both deterministic and non-deterministic components.  

While TDC aims to enforce meaningful relationships between static and dynamic states in complex dynamical systems, we hypothesize that a more refined representation may be necessary to capture both deterministic and non-deterministic components. Deterministic nodes are organized as static and dynamic pairs, constrained by the TDC loss to model physically meaningful variables. In contrast, statistical nodes are unconstrained and are used to represent residual or discrete behaviors such as pump activation or valve state, which do not follow physical dynamics.

\subsection{Dataset}
The evaluation is conducted using the publicly available BATADAL dataset, which stands for BATtle of the Attack Detection ALgorithms \cite{taormina_battle_2018}. This dataset was developed as part of a challenge aimed at comparing the performance of algorithms designed to detect cyber-attacks in water supply systems. It represents the C-Town water distribution system, modeled after a real-world, medium-sized network. The system includes 429 pipes, 388 junctions, 7 storage tanks ($T1-T7$), 11 pumps ($PU1-PU11$) distributed across 5 pumping stations, 5 valves ($V1-V4$), a single reservoir, and nine Programmable Logic Controllers (PLCs). The dataset contains 43 numerical metrics recorded on an hourly basis, with each feature vector representing one hour and consisting of 43 features. These features are readings from the SCADA system, including 7 features representing tank water levels, 12 features related to inlet and outlet pressures for an actuated valve and all pumping stations, and 24 features representing their flow rates and operational status.

We adopt the dataset split from this study \cite{somma_anomaly_2025}, where the BATADAL dataset is divided into three segments, each representing a distinct edge area within the water distribution network. This segmentation is based on the network's topology, ensuring that the data for each model is closely aligned with the physical layout and operational relevance of the Programmable Logic Controllers (PLCs) situated in these areas. Tab.~\ref{tab:EdgeFeatures} presents the features used for each segment, illustrating how the dataset has been tailored to reflect the operational structure of the system.
\begin{table}[h]
\centering
\caption{Features of the BATADAL Dataset corresponding to different edge areas based on \cite{somma_edge-based_2024}.}
\label{tab:EdgeFeatures}
\begin{tabular}{c|p{6cm}} 
\toprule
\textbf{Edge Areas} & \textbf{Dataset Features} \\
\midrule
Edge 1 & L\_T1, F\_PU1, S\_PU1, F\_PU2, S\_PU2, F\_PU3, S\_PU3, P\_J280, P\_J269 \\ \midrule
Edge 2 & L\_T2, L\_T3, L\_T4, F\_PU4, S\_PU4, F\_PU5, S\_PU5, P\_J300, P\_J256, F\_PU6, S\_PU6, F\_PU7, S\_PU7, P\_J289, P\_J415, P\_J14, P\_J422, F\_V2, S\_V2 \\\midrule
Edge 3 & L\_T5, L\_T6, L\_T7, F\_PU8, S\_PU8, F\_PU9, S\_PU9, P\_J302, P\_J306, F\_PU10, S\_PU10, F\_PU11, S\_PU11, P\_J307, P\_J317 \\
\bottomrule
\end{tabular}
\end{table}

\subsection{Performance Metrics}

 The most common performance metrics for anomaly detection applications are used, including confusion matrix and its derivatives -- $F1$ score, true positive rate ($TPR$), true negative rate ($TNR$) and positive predictive value ($PPV$).

In order to compare results to literature, some additional performance metrics proposed specifically for the BATADAL challenge \cite{taormina_battle_2018} were adopted. The performance score \( S_{TTD} \) reflects the time-to-detection, measuring the interval between the start of an attack and its detection, where \( S_{TTD} = 1 \) indicates immediate detection and \( S_{TTD} = 0 \) indicates no detection. The classification performance score \( S_{CLF} \) is defined as the mean of the true positive rate (TPR) and true negative rate (TNR). Finally, the overall ranking score \( S \) is computed as the mean of \( S_{TTD} \) and \( S_{CLF} \).

\subsection{Data Preprocessing}
To ensure our models handle the input data effectively, it is necessary to normalize the features. While standard normalization—subtracting the mean and dividing by the standard deviation—is a typical choice, we employ robust scaling due to its proven effectiveness in previous studies \cite{stojanovic_enhanced_2022,somma_edge-based_2024}. Robust scaling relies on the median and the interquartile range (IQR), which is the difference between the 75th and 25th percentiles. This approach is more resistant to outliers, preventing extreme values from having an undue influence on the feature distribution. As a result, the scaled features tend to have a standard deviation near 1, without being distorted by the presence of outliers.
\section{Temporal Differential Consistency Informed Autoencoders in Water Distribution Networks}

The motivation behind this approach is to embed physics-inspired consistency principles into the autoencoder while extending its applicability beyond purely data-driven representations. Traditional autoencoders primarily encode statistical features, which may not fully capture the structured temporal dependencies present in many physical and engineered systems. Inspired by \cite{somma_anomaly_2025} and the Lagrangian formalism \cite{goldstein_classical_2002}, we explore whether integrating these ideas into a hybrid autoencoder can enhance its ability to model a broader range of complex dynamical systems. Specifically, we hypothesize that by structuring the latent space with both dynamic and static nodes, the model can more effectively capture abrupt anomalies, such as tank overflows, in addition to gradual degradation processes like continuous wear \cite{somma_anomaly_2025}. Furthermore, we investigate whether this structured latent space allows the model to represent system evolution without explicitly using physical laws. Our goal is to assess whether this approach can be applied to diverse systems, such as water distribution networks, where both deterministic and stochastic effects influence system behavior.


\subsection{Structuring the Latent Space}
As a start we need a way to approximate the first time derivative. The \textit{central difference derivative} is a numerical method used to approximate the derivative of a function. It is based on evaluating the slope between two points symmetrically spaced around the point of interest. 

To implement this approach, we split the latent space into deterministic and statistical nodes. The deterministic nodes are organized in pairs, with each pair consisting of a static representation and its corresponding first-time derivative, thereby capturing both the static and dynamic aspects of the system.

In the case of the BATADAL data, this structure is static features may include elements like tank stand levels, which remain relatively constant over time, while dynamic features could encompass variables such as flow rates, which vary and reflect the system's evolving state. Additionally, there are features like valve states, which can be either open or closed. Such behaviors are difficult to model using physical deterministic laws due to challenges like exploding gradients. 

\subsection{Integrating Central Difference Calculation into the Latent Space}
 \begin{figure}[h!]
    \centering
    \includegraphics[width=0.8\columnwidth]{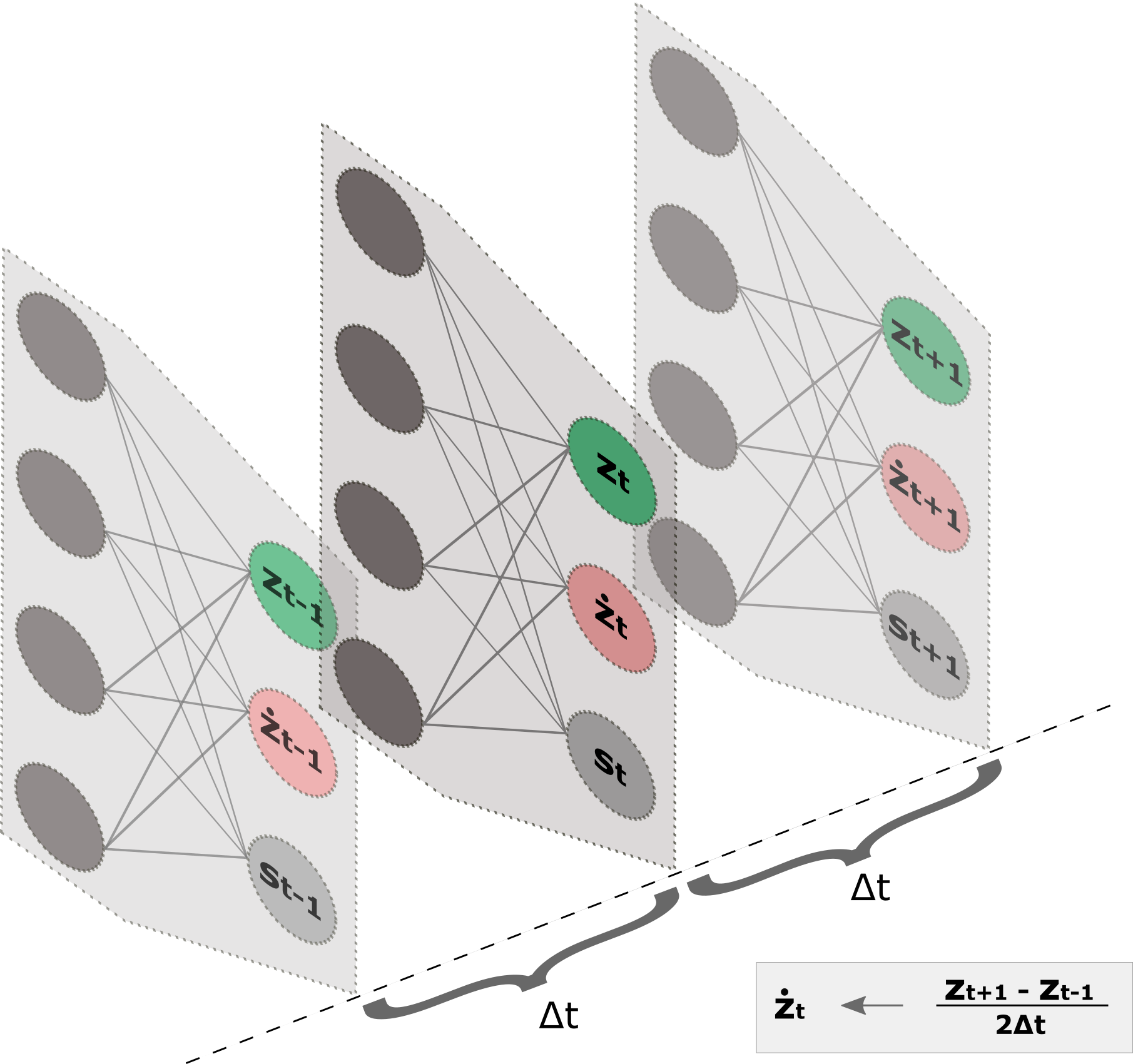}
    \caption{The encoder part of the autoencoder at the current, previous, and next time steps. The static deterministic nodes (green), their respective derivatives (red), and statistical nodes (grey) are shown schematically. We enforce the derivative nodes to reflect the central difference derivative.}
    \label{fig:TDC_scheme}
\end{figure}
In alignment with \cite{somma_anomaly_2025}, we compute the latent representations at the previous (\(t-1\)) and next (\(t+1\)) time steps and apply the central difference method to approximate the time derivative of the static nodes. This derivative is then used to ensure that the dynamic (derivative) nodes in the latent space approximate the central difference derivative of the respective static nodes, considering the time interval \(\Delta t\). This approach is shown schematically in Fig.~\ref{fig:TDC_scheme}. It is important to note that, since each feature is typically min-max scaled, this introduces an abstract time scale, which can potentially affect the interpretability of absolute derivative values.

\subsection{Informing Training Process with Temporal Differentials}
In this work, we build upon the TDC-AE \cite{somma_anomaly_2025}, which leverages the concept of \textit{temporal differential consistency loss (TDC-Loss)} to enforce structured temporal relationships in the latent space in complex dynamical systems. The TDC-Loss encourages the model to learn a representation that captures the system’s temporal evolution by ensuring that the time derivatives of latent deterministic nodes align with their estimated finite difference approximations. To extend this framework, we integrate conventional statistical latent nodes alongside TDC-informed latent nodes, leading to our proposed \textbf{hybrid Temporal Differential Consistency Autoencoder (hTDC-AE)}. A compact version of the pseudocode for hTDC-AE is presented in Algo.~1.


\begin{algorithm}
\label{alg:TDC_informed_training}
\caption{hTDC-AE: Hybrid Temporal Differential Consistency Autoencoder Training}
\begin{algorithmic}[1]
\STATE \textbf{Initialize:} Autoencoder with deterministic latent nodes as static/dynamic pairs (\(\mathbf{z}\), \(\dot{\mathbf{z}}\)), and statistical latent nodes \(\mathbf{s}\), Optimizer, MSE loss function, Weight \(\alpha\) TDC loss term.
\FOR{each epoch in training\_epochs}
    \FOR{each batch in training\_data}
        \STATE Perform forward pass through the Autoencoder and Encoder:  
        \STATE \(\quad \tilde{X}_t \gets \text{Autoencoder}(X_t)\)  
        \STATE \(\quad (\mathbf{z}_{t-1}, \dot{\mathbf{z}}_{t-1}, \mathbf{s}_{t-1}) \gets \text{Encoder}(X_{t-1})\)  
        \STATE \(\quad (\mathbf{z}_{t+1}, \dot{\mathbf{z}}_{t+1}, \mathbf{s}_{t+1}) \gets \text{Encoder}(X_{t+1})\)  
        \STATE Compute central difference derivative:  
        \STATE \(\quad \Delta_t \mathbf{z} \gets ( \mathbf{z}_{t+1} - \mathbf{z}_{t-1}) / 2\Delta_t\)  
        \STATE Compute temporal difference consistency loss using MSE:  
        \STATE \(\quad \text{TDC-Loss} \gets \text{MSE}(\Delta_t \mathbf{z}, \dot{\mathbf{z}_t})\)  
        \STATE Compute reconstruction loss using MSE:  
        \STATE \(\quad \text{Rec-Loss} \gets \text{MSE}(\tilde{X}_t, X_t)\)  
        \STATE Compute total loss:  
        \STATE \(\quad \mathcal{L} = \text{Rec-Loss} + \alpha \cdot \text{TDC-Loss}\)  
        \STATE Backpropagate the total loss: compute gradients w.r.t. model parameters  
        \STATE Update the Autoencoder parameters using the optimizer  
    \ENDFOR
\ENDFOR
\end{algorithmic}
\end{algorithm}
\subsection{Anomaly Estimation based on Autoencoders}  

Anomaly detection is carried out independently on the test dataset for each edge area using the trained Autoencoders. For each data sample, the reconstruction error is calculated, which forms the basis for detecting anomalies. To smooth out any fluctuations, a moving average filter with a window size of 7 is applied to the reconstruction errors, as suggested 
by \cite{stojanovic_enhanced_2022}. The threshold is defined as the 95th percentile of the reconstruction errors of the training data. Samples where the reconstruction error surpasses this threshold are classified as anomalies.

\subsection{Hardware and Implementation Details}
\begin{table}[h]
\centering
\caption{Detailed representation of the hyperparameters used in the training of the different Autoencoders.}
\label{tab:hypParameterList}
\begin{tabular}{p{3cm}p{1.4cm}p{1.4cm}p{1.4cm}}
    \toprule    
    \textbf{Hyperparam.} &  \textbf{Edge 1} & \textbf{Edge 2} & \textbf{Edge 3} \\
    \midrule
    optimizer & Adamax & Adamax & Adamax \\ 
    activation  & tanh & tanh & tanh \\ 
    no. dense layers  & 2 & 2 & 2 \\       
    no. neurons per layer  & 9 & 19 & 15 \\
    No. latent neurons (static-dynamic, statistic) & (3-3, 1) & (3-3, 2) &  (3-3, 2) \\ 
    learing rate & 0.01 & 0.007 & 0.01 \\ 
    batchsize & 32 & 32 & 32 \\
    $\alpha$ & 0.002 & 0.003 & 0.002 \\
    \bottomrule
\end{tabular}
\end{table}

Our experiments were conducted on a Windows 11 machine equipped with a 13th Gen Intel Core i7-13700H processor (2.4 GHz, 14 cores, 20 threads), 32 GB of RAM, and an NVIDIA GeForce RTX 4070 Laptop GPU. 

For implementation, we utilized PyTorch \textit{2.3.0} with CUDA \textit{11.8} for GPU-accelerated training. The autoencoder  architecture and training parameters are summarized in Tab.~\ref{tab:hypParameterList}. The model was trained using GPU acceleration to optimize performance and reduce training time. The code is available upon request.

\section{Results}
\subsection{The Temporal Differential Consistency Training}
 \begin{figure}[!h]
    \centering
    \includegraphics[width=0.9\columnwidth]{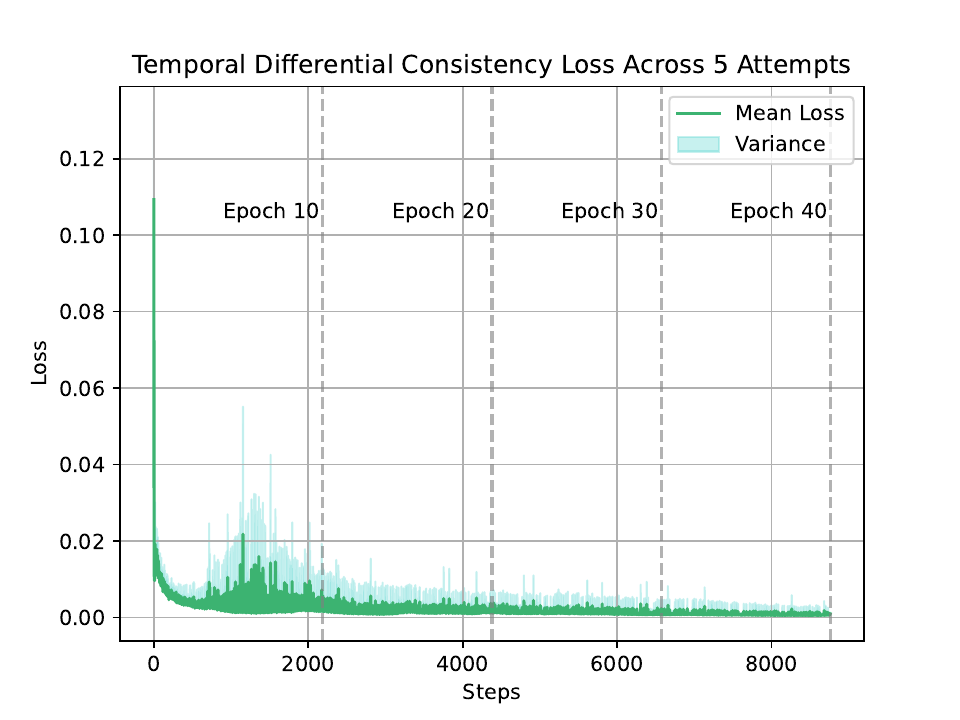}
    \caption{Temporal Differential Conisitency loss term evolution over 40 Epochs for 5 independent trainings.}
    \label{fig:temporalConsistencyLoss}
\end{figure}
In Fig. \ref{fig:temporalConsistencyLoss}, we present the temporal differential loss term across five independent training attempts over 40 epochs. The plot shows a clear pattern of convergence and reproducibility, indicating that the model consistently learns the underlying temporal features of the data.

 \begin{figure}[h]
    \centering
    \includegraphics[width=0.85\columnwidth]{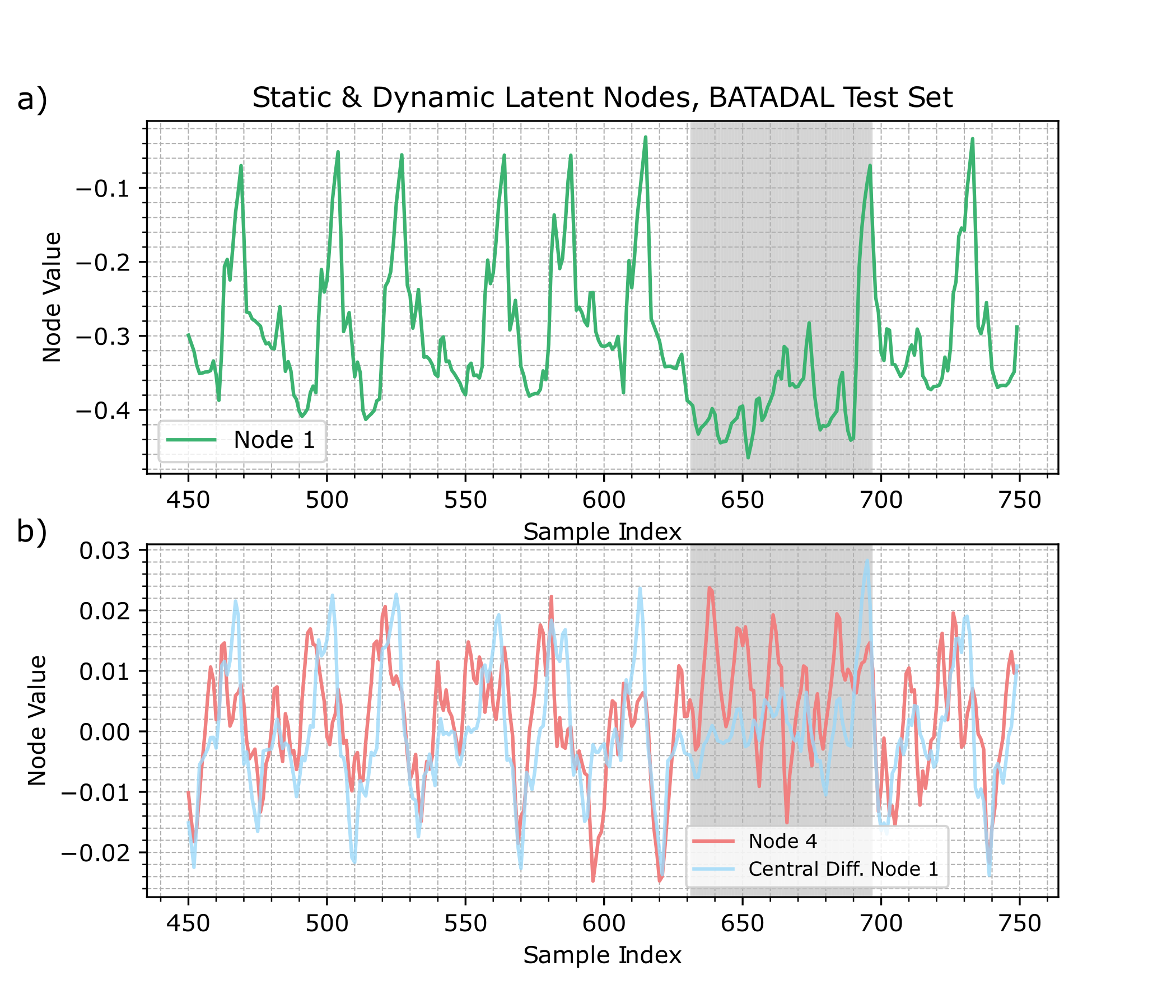}
    \caption{Example of a deterministic latent node pair (static/dynamic) on 300 samples from the BATADAL test dataset, including one attack phase (highlighted in grey).}
    \label{fig:DeterministicLatentNodes}
\end{figure}

To further investigate the latent space learned by our autoencoder, we focus on a pair of deterministic latent nodes (nodes 1 and 4) and plot their values across 300 test samples. This test set was not seen during training. To reduce visual noise, we applied a moving average with a window size of 2. In Fig.~\ref{fig:DeterministicLatentNodes}a the resulting traces of node 1 exhibit a clear periodic structure, which aligns well with the known underlying periodicity in the physical processes of the water distribution network. It appears to capture an aggregate of multiple periodic signals of varying frequencies and magnitudes, reflecting how different system features contribute to the overall dynamic.
Next, we compare dynamic node 4 to the central difference of node 1 to assess whether node 4 effectively encodes the temporal evolution of the state captured by node 1. As shown in Fig.~\ref{fig:DeterministicLatentNodes}b, the alignment between the two signals is reasonably close but not perfect. We hypothesize that the observed misalignment between the central difference of node 1 and the trajectory of node 4 arises from two key challenges. First, the comparison itself is inherently difficult due to the challenges of numerically differentiating noisy signals is an ill-posed problem. It requires smoothing and filtering techniques \cite{van_breugel_numerical_2020}. In this case, an averaging window of 7 timesteps was applied. Second, the complexity of the underlying system, combined with its partial observability, raises fundamental challenges in deriving an exact representation of the system’s dynamics from the available measurements. 
In a previous study \cite{somma_anomaly_2025}, the theoretical conditions necessary to reconstruct the system’s dynamics from measured values and the corresponding dimensional requirements were examined. However, when discrete measurement values, such as valve positions or pump activation states, are introduced, it is no longer a trivial consideration under which conditions an exact description of the system’s dynamics can be formulated. Despite these challenges, for the purposes of our use case, we don't aim to have an exact derivative. Node 4 gives an indication whether the state captured by node 1 is increasing or decreasing, providing an approximate causal link between the state and its rate of change. Crucially, this approximate link consistently breaks down during attack phases, as observed across all attack scenarios in the BATADAL test dataset. On example can be observed in detail in Fig.~\ref{fig:DeterministicLatentNodes}b. It is this disruption in the causal link within the dynamic description of the system that we leverage to enhance both the detection capability and the response time of anomaly detection in complex operational systems.

\begin{figure}
    \centering
    \includegraphics[width=1.\columnwidth]{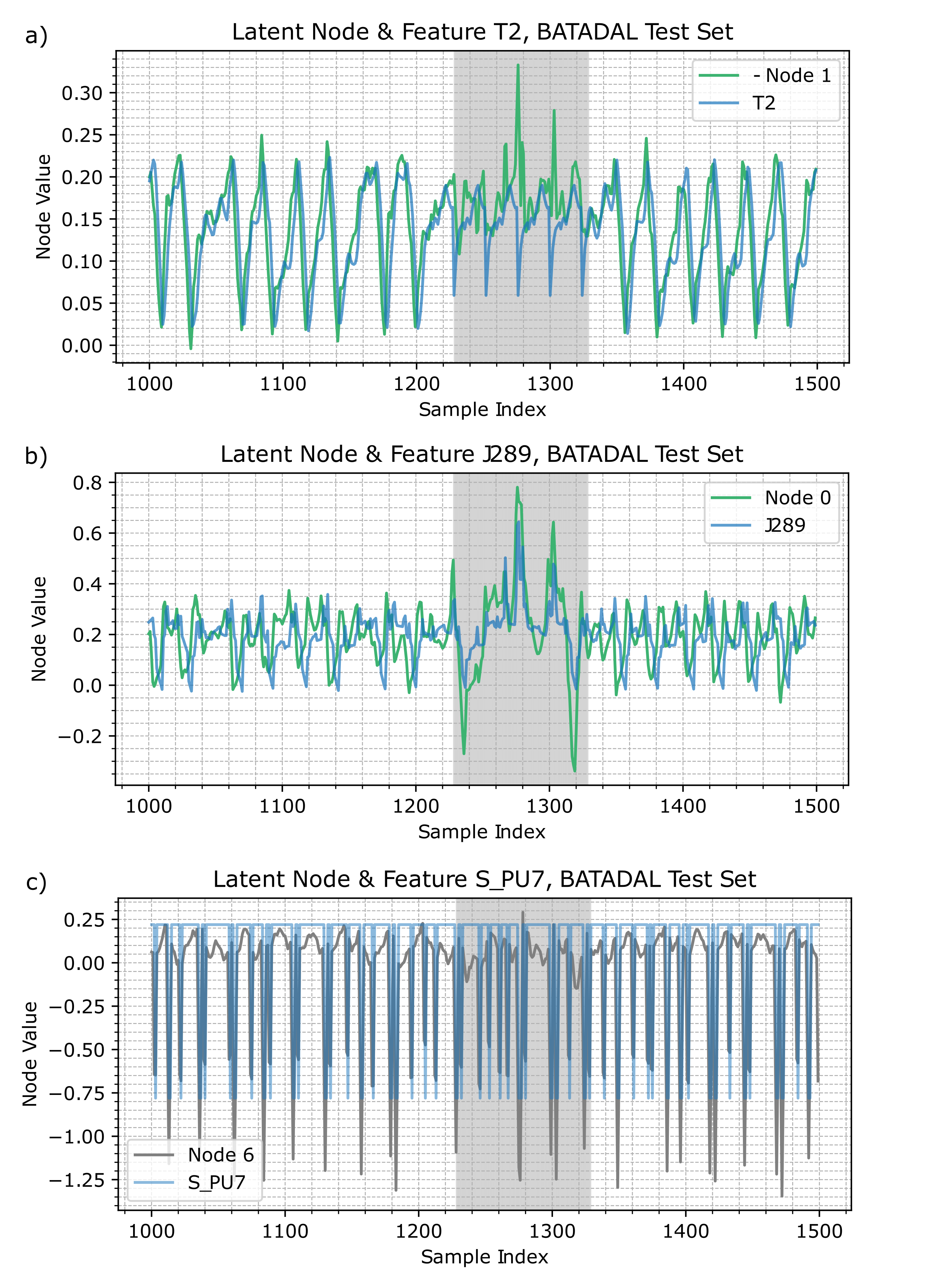}
    \caption{Comparison of three different input features and their highly correlated latent nodes across 500 examples from the BATADAL test dataset, including one attack phase (highlighted in grey). The input features were scaled, and the offset was adjusted to facilitate the comparison.}
    \label{fig:Feature_Nodes_Corr}
\end{figure}

Next, we analyze the relationship between input features and their latent space representations. For clarity, the scale and y-offset of the latent nodes were adjusted to align with the input features. In Fig.~\ref{fig:Feature_Nodes_Corr}a, the time series of a tank feature (T2) is overlaid with latent node 1, revealing a strong correlation. The negative sign of the correlation arises from the tanh activation function's symmetry and random weight initialization but still encodes the temporal dynamics. During the attack phase (T2 overflow), this relationship breaks down, and the tank feature exhibits variability far beyond normal operational ranges, highlighting the anomaly.

In Fig.~\ref{fig:Feature_Nodes_Corr}b, a flow value at a junction (J298) is compared to its corresponding latent node. While the correlation remains strong, the latent representation incorporates contributions from other features, reflecting the system's interconnected nature. As with T2, the attack phase disrupts this correlation, underscoring the latent representation's sensitivity to anomalies.

Finally, Fig.~\ref{fig:Feature_Nodes_Corr}c shows latent node 6, a statistical node capturing non-deterministic features like pump activations. The alignment validate our hypothesis that statistical nodes encode abrupt, discrete behaviors.

These results suggest that the model starts to distinguish and meaningfully represent both deterministic and statistical features, indicating a potential step towards more interpretable latent representations for anomaly detection in complex systems.

\subsection{Detection Performance}
To assess the overall performance impact of our training approach, Tab. \ref{tab:res} compares the classification performance metrics against state-of-the-art references. Our model achieves a performance of 0.9555, closely matching the AE-based benchmark of 0.9587 \cite{stojanovic_enhanced_2022}. The leading approach in the BATADAL challenge \cite{housh_detecting_2022} performs slightly better but relies on intricate physical modeling of the water distribution process. While effective, this method requires substantial domain expertise, making it less practical for real-world applications.

In a previous study \cite{somma_edge-based_2024}, it was shown that an edge-based approach could improve the mean time to detect anomalies. With the introduction of the proposed algorithm, we further enhance this detection time by an additional 3\%. This outperform the current benchmarks \cite{somma_edge-based_2024,housh_model-based_2018}.  Notably, our model reduces the number of fully connected layers from three to two, which not only simplifies the architecture but also decreases computational requirements. This reduction aligns with our broader goal of developing greener intrusion detection systems and contribute to the overarching societal goals of sustainability. Despite its simplicity and minimal reliance on domain-specific knowledge, our model slightly outperforms the challenge leader, offering a more generalizable and accessible solution for anomaly detection in water distribution systems.

\begin{table*}[h]
    \caption{A comparison of the performance of the proposed algorithm and algorithms from literature on the BATADAL dataset.}
    \label{tab:res}
    \centering
    \begin{tabular}{llllllllllllll}
        \emph{No.} & \emph{Reference} & \emph{Approach} & $S$ & $S_{TTD}$ & $S_{CLF}$ & $F1$ & $TPR$ & $TNR$ & $PPV$ & $TP$ & $FP$ & $TN$ & $FN$ \\
        \midrule
         \textbf{1} & \textbf{Proposed} & \textbf{hTDC-AE} & \textbf{0.9765} & \textbf{0.9974} & \textbf{0.9555} & \textbf{0.9176} & \textbf{0.9165} & \textbf{0.9944} & \textbf{0.9187} & \textbf{-} & \textbf{-} & \textbf{-} & \textbf{-} \\
        2 & \cite{housh_model-based_2018,taormina_battle_2018} & B1 & 0.9701 & 0.9650 & 0.9752 & 0.9700 & 0.9533 & 0.9970 & 0.9873 & 388 & 5 & 1677 & 19\\
        3 & \cite{stojanovic_enhanced_2022} & Enh. AE & 0.9571 & 0.9556 & 0.9587 & 0.9167 & 0.9459 & 0.9714 & 0.8891 & 385 & 48& 1628 & 22 \\
        4 & \cite{somma_edge-based_2024} & Edge AE & 0.9534 & 0.9701 & 0.9469 & 0.9253 & 0.9056 & 0.9913 & 0.9215 & - & - & - & - \\
        5 & \cite{ramotsoela_attack_2019} & QDA & 0.9495 & 0.9584 & 0.9406 & 0.8981 & 0.9091 & 0.9721 & 0.8873 & 370 & 47 & 1635 & 37 \\
        6 & \cite{abokifa_detection_2017,taormina_battle_2018} & B2 & 0.9491 & 0.9580 & 0.9402 & 0.8813 & 0.9214 & 0.9590 & 0.8446 & 375 & 69 & 1613 & 32 \\
        7 & \cite{giacomoni_identification_2017,taormina_battle_2018} & B3 & 0.9267 & 0.9360 & 0.9174 & 0.9057 & 0.8378 & 0.9970 & 0.9855 & 341 & 5 & 1677 & 66 \\
        8 & \cite{ramotsoela_attack_2019} & MD & 0.9165 & 0.9069 & 0.9260 & 0.8920 & 0.8722 & 0.9798 & 0.9126 & 355 & 34 & 1648 & 52 \\
        9 & \cite{ramotsoela_attack_2019} & Ensemble & 0.9142 & 0.8998 & 0.9286 & 0.8856 & 0.8845 & 0.9727 & 0.8867 & 360 & 46 & 1636 & 47 \\
        10 & \cite{choi_improvement_2020} & OS-ELM & 0.9100 & 0.9400 & 0.8800 & 0.8060 & 0.8080 & 0.9520 & 0.8040 & 329 & 80 & 1602 & 78 \\

        \bottomrule
    \end{tabular}
\end{table*}

\section{Discussion \& Outlook}
In this paper, we introduce \textbf{hTDC-AE}, an extension of TDC-AE~\cite{somma_anomaly_2025}. This hybrid approach enhances latent space structuring for anomaly detection by combining a vanilla autoencoder with the previously proposed temporal differential consistency (TDC) loss~\cite{somma_anomaly_2025}.
Our approach demonstrates stable and reproducible results across multiple training runs. By organizing the latent space into deterministic and statistical nodes, the model shows the potential to bridge the gap between data-driven methods and the integration of physics-inspired principles. Deterministic nodes show indications of encoding features related to physical processes, while statistical nodes capture non-deterministic, abrupt changes such as pump activations. Furthermore, our results suggest that a TDC-based approach enhances the model’s ability to detect not only gradual deviations but also more sudden anomalies, such as tank overflows or abrupt pressure drops, making it applicable to a wider range of real-world scenarios. Dynamic nodes exhibit strong temporal correlations, approximating the causal link between system states and their derivatives. We attribute the enhanced detection capability to the breakdown of this approximated causal link, which translates into a faster increase in reconstruction error, enabling quicker and more reliable anomaly identification. Our approach matches the classification performance of state-of-the-art models while significantly improving the \textbf{time to detect anomalies}, a key performance metric in real-world cyber-physical systems, by 3\%. This improvement is critical for enhancing system responsiveness in practice. Notably, our model outperforms the leader of the BATADAL challenge without relying on specific domain knowledge, making it a more generalizable solution. Furthermore, our method maintains the same time complexity as conventional autoencoders while reducing the number of fully connected layers from three to two compared to the benchmark. This simplification not only streamlines the architecture but also contributes to a \textbf{greener, more sustainable model} with lower computational requirements. Our algorithm is highly flexible and can be integrated into any autoencoder for anomaly detection in systems where physical data or deterministic processes are expected. We aim to move beyond conventional statistical fitting of non-linear functions, contributing to the broader challenge of guiding neural networks to learn \textbf{generalizable and interpretable abstractions} that embed meaningful, process-driven relationships.

Future work will focus on studying the \textbf{theoretical foundations} of this method and testing it on simpler systems with known analytical solutions. This will allow us to validate the relationships between static and dynamic features in the latent space and explore whether consistent laws emerge between deterministic nodes. Such investigations could provide deeper insights into the system’s dynamics and further study the interpretability of the TDC-informed models.

\section*{Acknowledgements}
This work was funded by the European Union's Horizon Research and Innovation Programme under GA No. 101119681 (project ResilMesh) and by the Austrian Security Research Programme KIRAS of the Federal Ministry of Finance (BMF) under GA No. FO999905309 (project SeRWas)

\input{ref.bbl}

\end{document}